\documentclass[reprint,
twocolumn,
amsmath,amssymb,
aps,pre
]{revtex4-2}

\usepackage{graphicx}
\usepackage{dcolumn}
\usepackage{bm}

\newcommand{\enquote}[1]{``#1''}

\usepackage{color}

\begin{document}


\title{Local elastic perturbation of colloidal suspensions near the colloidal glass transition}

\author{Piotr Habdas}
\email{phabdas@sju.edu}
\affiliation{Department of Physics, Saint Joseph's University, Philadelphia, PA 19131}

\author{Rachel E. Courtland}
\altaffiliation{Current address:  MIT Technology Review, Cambridge, MA}
\affiliation{Department of Physics, Emory University, Atlanta, GA 30322}

\author{Eric R. Weeks}
\email{erweeks@emory.edu}
\affiliation{Department of Physics, Emory University, Atlanta, GA 30322}

\date{\today}

\begin{abstract}
Isolated microscopic magnetic particles are used to induce local perturbations in dense colloidal suspensions by rotating an external magnet. Confocal microscopy enables tracking of both the magnetic probe particle and adjacent colloidal particles.  A probe particle moves with a circular trajectory.  Knowing the external force and measuring the amplitude and phase of the probe motion allows us to infer the elastic and viscous moduli of colloidal suspensions at various volume fractions. These measurements are in qualitative agreement with previous results from conventional rheology. To further analyze the system’s response, the oscillatory amplitude of colloidal particles is evaluated as a function of distance from the probe, revealing a $1/r$ decay in amplitude, consistent with a homogeneous viscoelastic material. These observations confirm that continuum descriptions of the colloidal samples are effective down to length scales comparable to the particle diameter.
\end{abstract}

\maketitle

\section{Introduction}
Colloidal suspensions near the glass transition exhibit a pronounced evolution in structure and mechanical response, changing from viscous, fluid-like behavior to dynamically arrested, solid-like states as the particle volume fraction $\phi$ increases \cite{pusey86,pusey87,vanmegen91,vanmegen93,vanmegen94,mason95,bergenholtz99,trappe01,pham02,schweizer03,hunter12rpp}. This transition is accompanied by slow structural relaxation, particle caging, and spatially heterogeneous dynamics, making dense suspensions a model system for studying glassy behavior in disordered materials \cite{pusey86,pusey87,vanmegen91,vanmegen93,vanmegen94,segre95,vanmegen98,kasper98,schweizer03,hunter12rpp}. A central issue is how microscopic particle-scale interactions give rise to macroscopic rigidity and viscoelastic response not only in hard-sphere colloidal suspensions, but also in jammed soft-particle systems \cite{khabaz20}.

In dense suspensions of nearly hard colloidal spheres, where interparticle interactions are dominated by excluded-volume constraints, the equilibrium free energy is predominantly entropic and arises from the multiplicity of accessible particle configurations \cite{carnahan69,pusey86}. As the colloidal glass transition is approached, particles become increasingly caged by their neighbors, dramatically slowing structural relaxation and producing solid-like behavior on intermediate time scales \cite{bengtzelius84,pusey86,gotze91,vanmegen93}. Under small-amplitude shear, deformation distorts the otherwise isotropic pair structure into an anisotropic configuration that is statistically less probable, thereby reducing configurational entropy and increasing the free energy. Because relaxation is slow near the glass transition, this entropy reduction cannot be rapidly recovered, and the stored free energy manifests as a finite elastic stress and a storage modulus that grows strongly with volume fraction \cite{brady93,mason95,fuchs02,swan14,liu22}. The viscoelastic response therefore reflects a competition between entropic restoring forces associated with cage distortion and viscous dissipation due to structural rearrangements. As the glass transition is approached, the relaxation time diverges and the system exhibits an extended plateau in the elastic modulus, characteristic of an arrested, entropically dominated microstructure \cite{vanmegen93,mason95,pham02}.

Mechanical properties near the colloidal glass transition are most commonly measured using bulk rheology, which provides the elastic (storage) and viscous (loss) moduli, $G'(\omega)$ and $G''(\omega)$.  Bulk rheology averages over the sample volume, and thus, such measurements do not directly reveal how mechanical perturbations manifest at the scale of the disordered microstructure. Confocal microscopy, in contrast, provides particle-resolved trajectories and has been widely used to characterize heterogeneous dynamics and cage rearrangements \cite{vanblaaderen95,kegel00,weeks00,ghosh11,jensen14,lu16,heckendorf17,roller21,ilhan22,singh23}. Extending imaging techniques to situations where a controlled localized force is applied enables direct measurement of the spatial structure of the mechanical response \cite{lee08,habdas04,anderson13,habdas25}.

Magnetic probe particles provide a means to impose localized forcing without mechanical contact. When driven by external fields, these probes act as embedded force centers whose motion can be controlled in amplitude and frequency. Driven probes have previously been used to study forced motion and microrheology near the colloidal glass transition \cite{habdas04,mason06,gazuz09,anderson13,gruber16,habdas25}. Here we focus on the linear oscillatory response of a magnetic probe and on the resulting displacement field of surrounding colloidal particles.

In a linear viscoelastic continuum, the far-field displacement induced by a localized oscillatory force decays as $1/r$, where $r$ is the distance from the forcing center \cite{palierne90,crocker00,levine00,levine01}. Measuring this spatial dependence provides a direct test of the applicability of continuum descriptions at length scales comparable to only a few particle diameters. At the same time, the probe amplitude and phase lag relative to the applied force provide access to $G'(\omega)$ and $G''(\omega)$ through generalized Stokes relations \cite{ziemann94,mason97,waigh05}.

In this work, we drive microscopic magnetic particles along circular trajectories in dense colloidal suspensions with volume fractions approaching the glass transition, while using confocal microscopy to track both the probe and surrounding colloidal particles. From the probe motion we determine $G'(\omega)$ and $G''(\omega)$, and from particle-resolved trajectories we observe $1/r$ spatial decay of the mean induced displacement field. This matches the expectation for a homogeneous continuum viscoelastic material noted above, even for distances $r$ comparable to the particle diameter.  Our results confirm that averaging over the response of particles near the magnetic probe, these dense colloidal samples behave in many ways as a continuum even down to the particle scale.

\section{Experimental Methods}
The colloidal particles in this study are sterically stabilized poly-(methylmethacrylate) (PMMA) spheres (radius $a=1.1$ $\mu$m) containing rhodamine dye.  The polydispersity in particle size is $\sim$5\%, which slows crystallization.  The PMMA spheres are suspended in a mixture of 85\% cyclohexylbromide and 15\% decalin by weight, which matches the density and index of refraction of the colloidal particles with the solvent.  The viscosity of the solvent is 2.18 mPa s at $22^{\circ}$C.

We add superparamagnetic particles with a radius of $a_{\rm MP}=2.25$~$\mu$m (M450, coated with glycidyl ether reactive groups, Dynal). The volume fraction of the magnetic probes is very small to prevent magnetic interaction between probe particles when in the presence of an external magnetic field. We do not observe attraction or repulsion between the colloidal particles and the magnetic beads, in either dilute or concentrated samples.  

In order to drive the magnetic probe particles, we pull them using a strong permanent magnet (neodymium, rare earth). The magnet is mounted on a steel caliper to allow the distance from the objective (and thus the force) to be adjusted.  To calibrate the forces, we separately use the magnet to pull a magnetic particle through glycerol.  The high viscosity of glycerol ($\eta = 1.5$~Pa$\cdot$s) results in a slow magnetic particle velocity, and using the Stokes drag law allows us to determine the exerted force which depends on the precise position of the magnet.  The forces we use range from 30.7 to 41.3~pN.

To quantify how fast a system is being forced with respect to its ability to dissipate the energy being inputted to the system, we define two ``modified'' P\'eclet numbers.   The first modified P\'eclet number, $Pe^*_{\rm osc}$, is a ratio of the time scale it takes for a colloidal particle to diffuse its own radius to the time scale of the oscillation of the magnetic particle:
\begin{equation}
    Pe^*_{\rm osc} =\frac {\tau_{\rm particles}}{\tau_{\rm osc}} = \frac{a^2/2D_\infty}{2\pi/\omega},
\end{equation} where $a$ is the colloidal particle radius, $\omega$ is the driving frequency, and $D_\infty$ is the long-time diffusion constant of the colloidal particles determined from the slope of the mean square displacement at long lag times.  It is this use of the $\phi$-dependent diffusivity that is the reason we term this the ``modified'' P\'eclet number (as opposed to using $D_0$, the diffusion constant for a dilute sample).  $D_\infty$ decreases as the colloidal glass transition is approached due to increased particle crowding \cite{vanmegen91,weeks00}.

The quantity $Pe^*_{\rm osc}$ alone does not provide complete information of how much the system is being perturbed, as the motion depends not only on $\omega$ but also on the amplitude $A_{\rm MP}$ of the magnetic particle's motion.  Hence, we define an additional modified P\'eclet number based on the magnetic particle displacement, 
\begin{equation}
    Pe^*_{\rm disp} =\frac {\tau_{\rm particles}}{\tau_{\rm disp}} = \frac{a^2/2D_\infty}{a/v},
\end{equation} where $v$ is the magnetic particle velocity.  Given that $v=A_{\rm MP} \omega$, these two P\'eclet numbers are related by $\mathrm{Pe}^*_{\rm disp} = [A_{\rm MP}/(2 \pi a)]\mathrm{Pe}^*_{\rm osc}$.  The term in square brackets is essentially a third nondimensional parameter which in our experiments ranges from 0.043 to 0.37.  The low end corresponds to $\phi=0.56$ and the highest forcing frequency, and the high end corresponds to $\phi = 0.32$ and the lowest forcing frequency.

In this research, $Pe^*_{\rm osc}$ and $Pe^*_{\rm disp}$ are both greater than one.  In the most dilute sample with $\phi$ = 0.32, 31 $<$ $Pe^*_{\rm osc}$ $<$ 94 and 354 $<$ $Pe^*_{\rm disp}$ $<$ 503.  In the most dense sample, $\phi=0.57$, $120 < \mathrm{Pe}^*_{\rm osc}< 370$ and $380 < \mathrm{Pe}^*_{\rm disp}< 642$.  Particle rearrangements are not observed in this regime; that is, all observed particle motion, both background particles and magnetic particles, is oscillatory.  Note that the range of the P\'eclet numbers is rather moderate in this work; our goal is not to focus on Pe dependence but rather that we stay in the Pe $> 1$ limit.  The key point is that we are perturbing the sample more rapidly than the relaxation time scale, and thus the colloidal particles do not undergo spontaneous rearrangements during the observations, but rather respond directly to the magnetic probe.

Dense colloidal samples are prepared through centrifugation in glass vials.  Next, the samples are diluted using the solvent mixture described above which matches the density and index of refraction of the colloidal particles.  Finally, diluted colloidal suspensions are transferred into microscopy chambers.  The volume fraction $\phi$ is determined by collecting 3D images using confocal microscopy.  Due to particle size uncertainty, this method is subject to a systematic uncertainty of about $\Delta \phi = \pm 0.02$, but our relative volume fractions are accurate to $\pm 0.002$ \cite{poon12}.  In order to prevent crystallization, sample slides are sonicated and placed on a slide stirrer overnight before acquisition. To find magnetic probes, the samples are observed in bright-field microscopy with a large field of view before switching to confocal microscopy.

To capture images of the probe particle and the colloidal particles, we use an inverted microscope and a $100 \times$ oil objective (Numerical Aperture of 1.4). We collect a series of 2D images ($80 \times 64$~$\mu$m$^2$) in the bulk of the sample (with image plane located at least 30 $\mu$m from the coverslip) at maximum rate of 7.5 images/s. Standard tracking techniques are used to determine particle trajectories \cite{crocker96}.  The uncertainty in particle positions is $\pm 0.05$~$\mu$m for both colloidal particles and probe particles.

\begin{figure}
    \includegraphics[scale=0.7]{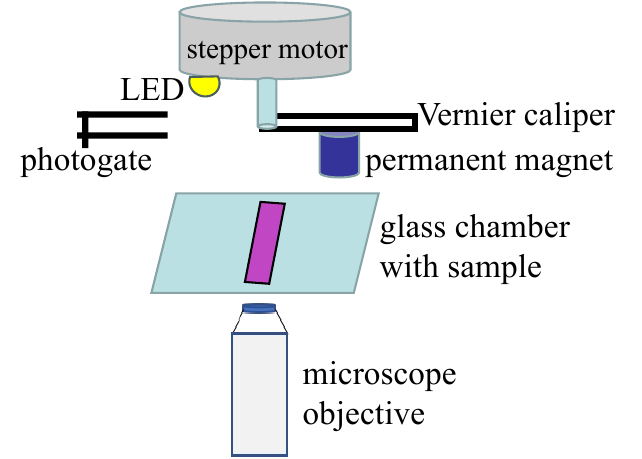}
    \caption{A schematic of the apparatus.}
    \label{apparatus}
\end{figure}

In order to create oscillatory forces, the caliper is mounted on the shaft of a NEMA-23 stepper motor [Intelligent Motor Systems (IMS)]. The motor sits on a three-axis micrometer stage, which is used to center the axis of rotation on the objective, as shown in Fig.~\ref{apparatus}. The motor is operated in constant angular velocity mode by controller software supplied by IMS though a serial connection. The motor is capable of speeds as high as 2 revolutions/s, and, since it has a microstep resolution of 51,200 steps per revolution, it is capable of low frequency motion without choppiness. 

To pinpoint the phase of the external magnet relative to the confocal movies, a photodiode is placed at a fixed position near the objective. At a certain point in each revolution, a flag placed on the rotating caliper interrupts the photogate, which momentarily turns on a red LED placed over the objective. The LED flash results in an overall increase in the intensity over the field of view. The intensity increase is typically about 15$\%$, lasts for a fraction of a second, and does not obscure the picture of the probe particles or the colloidal particles. The change in intensity is recorded by the confocal microscope and enables the later determination of the phase lag of the magnetic probe with respect to the external magnet.  In particular, we align the photodiode so that the rotating flag triggers the flash when the external magnet is aligned with the $y$-axis of the confocal image.  Given that the confocal frame rate is fixed and known, and the magnet rotation frequency is fixed and known, the data from the flashes throughout an entire experimental movie allow us to determine the absolute phase of the magnetic forcing to within $\pm 4^\circ$.  The phase lag of the magnetic probe is thus determined based on when the magnetic probe trajectory crosses the $y$-axis relative to the magnetic forcing phase.

The range of angular frequencies explored is limited by the onset of crystallization. Depending on the volume fraction, crystallization takes anywhere from 30 minutes to two hours to occur.  Data sets are typically 1200 frames long and 160 or 320 seconds in duration depending on the rotational frequency used. Frequency and volume fraction are the major variables in this experiment. Performing constant force measurements is fairly impractical, as the force is adjusted for different volume fractions to minimize the amount of out-of-plane motion and to obtain resolvable particle amplitudes without plastically deforming the sample. Occasionally, during the course of acquiring data, the magnetic particle moves out of plane of view so the focus was adjusted. No change in the dynamics was observed as a result of this migration. 

\section{Magnetic Particle Dynamics}

\begin{figure}
    \includegraphics[width=8cm,bb = 109 412 524 720]{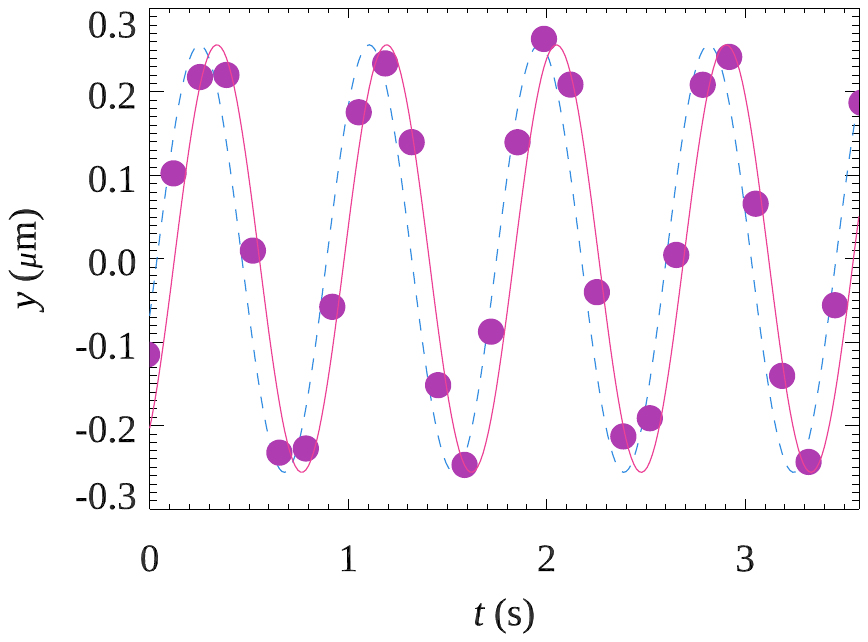}
    \caption{The symbols are the measured $y$ positions of the magnetic particle as a function of time.  The solid line is the sinusoidal fit to the position data using the total duration of the trajectory (160~s).  The dashed line is the magnetic forcing in the $y$ direction (arbitrary units) to demonstrate the phase lag.  For these data, $\omega = 3.68$~rad/s, $\phi = 0.56$, and the phase lag is $\psi_{\rm MP}=37^\circ$.}
    \label{ymagbead}
\end{figure}

\begin{figure}
    \includegraphics[width=8cm,bb = 42 392 442 760]{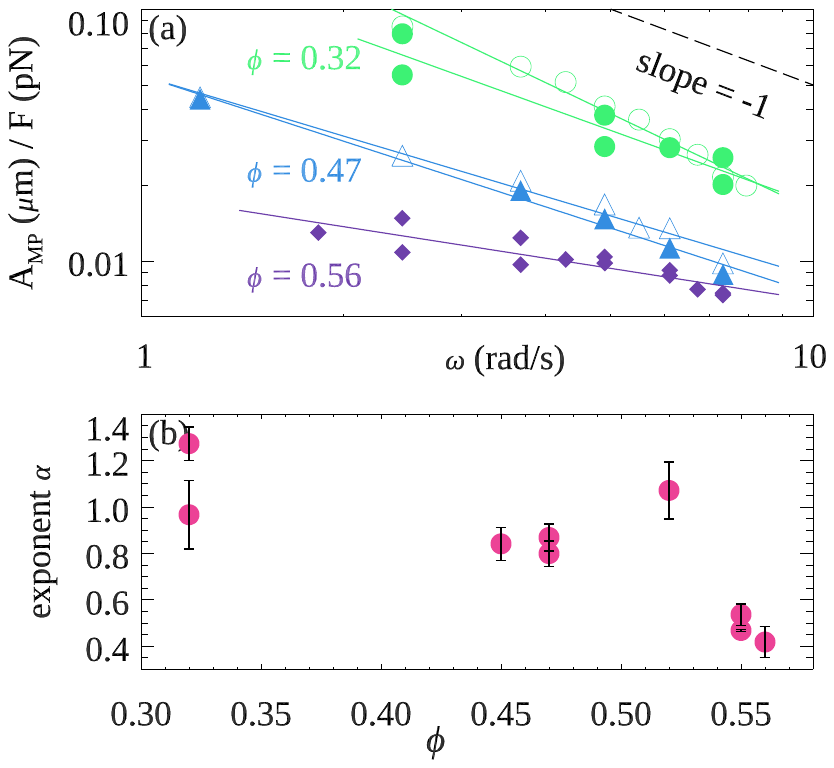}
    \caption{(a) The amplitude of the magnetic particle, $A_{\rm MP}$, scaled by the driving force, $F$, as a function of angular frequency for a selected volume fractions: $\phi =$ 0.32 (open and filled circles), 0.47 (open and filled triangles), and 0.56 (diamonds).  For clarity, data for $\phi =$ 0.45, 0.52, and 0.55 are not shown.  The solid lines are power-law fits to the single data sets. The dashed line in the top right corner is $A_{\rm MP}/F$ = 1/$\omega$.  Open and filled circles and triangles correspond to different experimental runs.   (b) The magnitude of the power-law fit exponent $\alpha$ from $A_{\rm MP}/F \sim \omega^{-\alpha}$ as a function of volume fraction $\phi$.  Error bars are uncertainties in the fit exponent from the fitting procedure.}
    \label{AoverFvsw}
\end{figure}

Using the procedures detailed above, we study how the colloidal suspension responds to a localized elastic perturbation as the colloidal glass transition is approached. We first focus on the dynamics of the magnetic particle and its response to the external forcing. From the trajectory of the particle, we readily extract two quantities: the oscillation amplitude of the magnetic particle, $A_{\rm MP}$ and its phase delay relative to the external forcing, $\psi_{\rm MP}$.  A short portion of a representative probe trajectory is shown in Fig.~\ref{ymagbead}.  We have three straightforward expectations for these quantities as the samples are varied from liquid-like to close to the glass transition, that is, as $\phi$ approaches $\phi_g \approx 0.58$.  First, the amplitude of the magnetic particle's motion will decrease as the sample's viscoelastic moduli increase with increasing $\phi$ (at fixed forcing).  Second, the phase lag will change from liquid-like ($\psi_{\rm MP} = \pi/2$) to solid-like ($\psi_{\rm MP} \rightarrow 0$).  That is, assuming the forcing looks like $F(t) = F \sin \omega t$, in a viscous sample the velocity is proportional to the forcing, so the position -- the integral of the velocity -- should be proportional as $x(t) \sim F/\omega \cos \omega t = F/\omega \sin (\omega t + \pi/2)$.  In an elastic solid, the position is directly proportional to the forcing, so then $x(t) \sim F (\sin \omega t + 0)$.  Third, these expectations also show $A_{\rm MP}$ should scale as $1/\omega$ for liquid-like samples and have no $\omega$ dependence for solid-like samples.  We have phrased these expectations in terms of linear motion, but the arguments hold true for circular motion as well.

We now show that all of these expectations are met in our data.  Given the expected linearity of the magnetic particle's amplitude on the force, we plot $A_{\rm MP}/F$ as a function of angular frequency $\omega$ in Fig.~\ref{AoverFvsw}(a) for several volume fractions.  As anticipated, $A_{\rm MP}/F$ decreases with increasing volume fraction over the frequency range studied. For dense samples, close to the colloidal glass transition, the motion of the magnetic particle is increasingly hindered, consistent with the growing elastic character of the suspension.

The solid lines in Fig.~\ref{AoverFvsw}(a) are power-law fits to the data, $A_{\rm MP}/F \sim \omega^{-\alpha}$.  Figure \ref{AoverFvsw}(b) presents the magnitude of the exponent $\alpha$ for the range of volume fractions studied.  As the volume fraction increases, the magnitude of the exponent decreases, matching the expectation of $\alpha \sim 1$ for liquid-like samples at low $\phi$ and $\alpha \rightarrow 0$ at higher $\phi$.

Note that the fit exponent for one of the data sets with $\phi$ = 0.32 [open circles in Fig.~\ref{AoverFvsw}(a)] has a power-law exponent greater than one. $\alpha > 1$ indicates that the suspension more strongly resists probe motion as frequency increases.  While this could indicate high-Pe shear-thickening behavior, this seems unlikely for a sample at this volume fraction  \cite{brady88,stickel05,vicic02,jones02,maranzano02,wagner09,cheng11,swan14}.  It is more likely that the probe particle is experiencing nonlinear effects at the lowest forcing frequency, which would result in strain-softening \cite{harrer12}.  Here, the magnetic force acts for a longer duration at any given position, allowing the force to distort the colloidal particles away from equilibrium \cite{gnann_asymptotic_2012,mohan14} while nonetheless not dragging the magnetic particle through the surrounding colloidal particles.  At higher frequencies, the nonlinear response is lessened, thus resulting in a misleading $\alpha > 1$.

\begin{figure}
    \includegraphics[width=8cm,bb = 119 412 532 720]{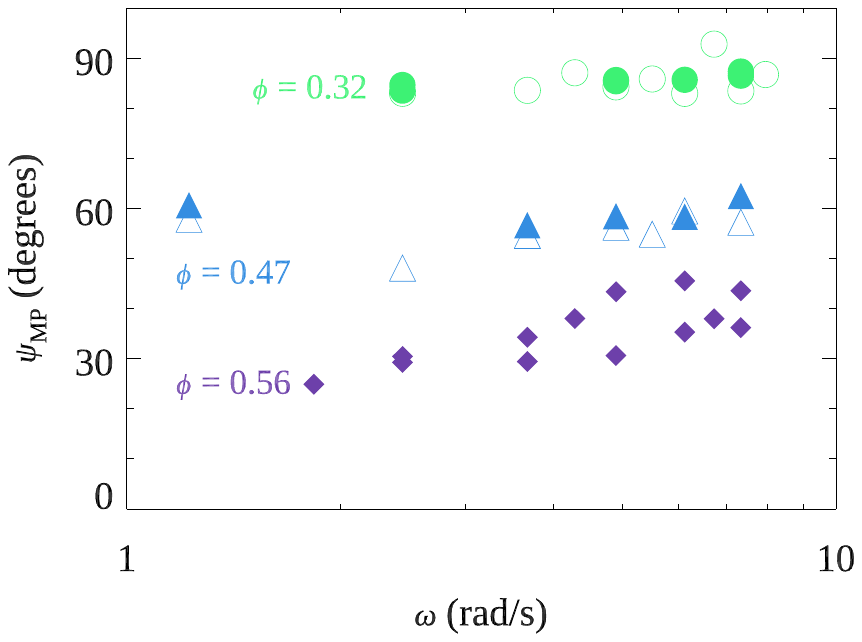}
    \caption{Phase lag of the magnetic particle, $\psi_{\rm MP}$, as a function of angular frequency for a selected volume fractions: $\phi =$ 0.32 (open and filled circles), 0.47 (open and filled triangles), and 0.56 (diamonds).  For clarity, data for $\phi =$ 0.45, 0.52, and 0.55 are not shown.}
    \label{PLvsw}
\end{figure}

The second physical quantity extracted from the position of the oscillating magnetic particle is the phase lag $\psi_{\rm MP}$, that is, where the probe position is relative to the magnetic forcing.  We plot $\psi_{\rm MP}$ as a function of angular frequency $\omega$ in Fig.~\ref{PLvsw}.  As expected, the phase lag is closer to $90^\circ$ for low volume fraction samples (more liquid-like) and decreases toward $0^{\circ}$ as the volume fraction of samples increases.

\begin{figure}
    \centering
    \includegraphics[width=8cm,bb = 60 364 368 737]{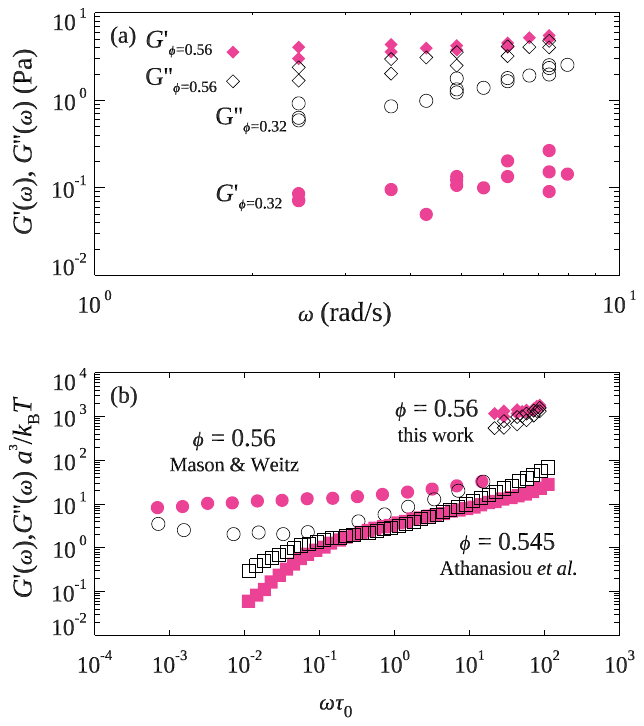}
    \caption{(a) Elastic modulus $G'(\omega)$ (red filled symbols) and viscous modulus $G''(\omega)$ (black open symbols) as a function of angular frequency for $\phi =$ 0.32 (circles) and 0.56 (diamonds). For clarity, data for $\phi =$ 0.32, 0.45, 0.47, 0.52, and 0.55 are not shown. (b) Scaled shear moduli, $G'(\omega)$ (red filled symbols) and $G''(\omega)$ (black open symbols), as a function of scaled angular frequency $\omega$ for data studied here, $\phi =$ 0.56 (diamonds), Mason and Weitz data (circles), $\phi =$ 0.56 \cite{mason95}, and Athanasiou {\it et al.} (squares), $\phi =$ 0.545 \cite{athanasiou25}.}
    \label{EMVMvsw2}
\end{figure}

The amplitude and phase of the magnetic particle can also be used to extract viscoelastic moduli via the methods of active microrheology \cite{gardel05}.  We can calculate the elastic (storage) modulus $G'$($\omega$) and the viscous (loss) modulus $G"$($\omega$) using:
\begin{equation}
\label{gprime}
G'(\omega) = \frac{F}{6\pi a_{\rm MP} A_{\rm MP}(\omega)} \cos \psi_{\rm MP}(\omega)
\end{equation}
and
\begin{equation}
\label{gdouble}
G''(\omega) = \frac{F}{6\pi a_{\rm MP} A_{\rm MP}(\omega)} \sin \psi_{\rm MP}(\omega)
\end{equation}
where $a_{\rm MP}$ is the magnetic particle radius, $A_{\rm MP}$ is the magnetic particle amplitude, and $\psi_{\rm MP}(\omega)$ is the phase lag of the magnetic particle with respect to the external forcing \cite{ziemann94}.  Given that we have shown $A_{\rm MP}$ and $\psi_{\rm MP}$ depend on the frequency $\omega$ and volume fraction $\phi$, Eqs.~\ref{gprime} and \ref{gdouble} make it clear that the moduli will also depend on these variables.  In particular, as $\psi_{\rm MP}$ changes from close to $90^\circ$ to close to $0^\circ$, the dominant modulus will change from the viscous modulus to the elastic modulus.  Figure~\ref{EMVMvsw2}(a) shows the frequency dependence of the viscoelastic moduli for suspensions at two selected volume fractions, $\phi = 0.32$ and $\phi = 0.56$. At the lower volume fraction, $\phi = 0.32$, $G'(\omega)$ (red filled circles) is of order $1\ \mathrm{Pa}$ while $G''(\omega)$ (black open circles) is of order $10\ \mathrm{Pa}$ over the accessible frequency range, indicative of fluid-like behavior consistent with the predominantly viscous behavior presented by the $\sim$ $90^{\circ}$ phase lag in Fig.~\ref{PLvsw}. Here, energy is primarily dissipated through viscous flow rather than stored elastically.

In contrast, at the highest volume fraction studied, $\phi = 0.56$, the moduli increase by more than an order of magnitude and the elastic modulus $G'(\omega)$ [red filled diamonds in Fig.~\ref{EMVMvsw2}(a)] becomes comparable to or larger than the viscous modulus $G''(\omega)$ [black open diamonds in Fig.~\ref{EMVMvsw2}(a)], consistent with the emergence of an elastic, solid-like response. This volume fraction dependent enhancement of $G'$ and the crossover between $G'$ and $G''$ with increasing $\phi$ highlight the progressive dynamical arrest of the suspension as it approaches the colloidal glass transition, where energy is primarily stored in the deformed particle configuration rather than dissipated.  
This plateau arises because at these volume fractions, particles are becoming increasingly trapped in cages formed by their neighbors, providing a quasi-permanent elastic network that responds on timescales much longer than our experimental frequencies.  This is because we are at high $Pe^*$:  the slow structural relaxation time scale does not matter at the faster forcing time scale we impose.  A similar weak frequency dependence and the emergence of a plateau in the elastic modulus have been reported in jammed soft particle glasses, where relaxation times become longer than experimental timescales \cite{liu22}.

The slight upturn in the viscous modulus $G''$ at higher frequencies [black open symbols, Fig.~\ref{EMVMvsw2}(a)] hints at the eventual high-frequency response.  Given that the colloidal sample has a liquid solvent with viscosity $\eta$, the eventual high-frequency response is expected to be $G'' \sim \eta \omega$ \cite{trappe01,caggioni20}. However, our frequency range is too narrow to see this crossover to the high-frequency asymptotic regime.

To facilitate comparison with rheological data in the literature, we scale frequency by $\tau_{0}$ ($=a^2/D_0$, with $D_0$ as the diffusion constant in the dilute limit), the microscopic relaxation time characteristic of the colloidal particles; the scaled result is shown in Fig.~\ref{EMVMvsw2}(b) for $\phi = 0.56$. Note that $\omega \tau_{0} = {\rm Pe}$, where Pe is a bare P\'eclet number which uses the bare diffusion constant, $D_0$, rather than the long-time diffusion constant $D_{\infty}$. Also, the shear moduli axis of Fig.~\ref{EMVMvsw2}(b) shows the reduced moduli scaled by $a^3/k_B T$ \cite{jones02}.  The graph also includes conventional rheology data from Refs.~\cite{mason95,athanasiou25}.  Our scaled elastic and viscous moduli are about an order of magnitude larger than those reported previously for these nominally similar colloidal suspensions. The similarity of the frequency dependence suggests that the same underlying viscoelastic mechanisms govern the response, while the difference in magnitude in the shear moduli likely reflects systematic differences between samples. The most probable origin for the modulus difference relates to the difficulty in comparing volume fractions between different experiments, with an uncertainty expected to be at least $\pm 0.01$ \cite{poon12}.  In particular, two factors matter.  First, our particles are known to have a slight electrostatic charge \cite{hernandez09,kurita10}, and this causes a larger effective hard-sphere volume fraction \cite{royall13,khabaz20}.  Second, our sample polydispersity is $\sim 5$\%, as compared to the prior experiments with polydispersity of 10\% \cite{athanasiou25} and 20\% \cite{mason95}.  Polydispersity shifts phase behavior to higher volume fractions \cite{royall13,sollich10}, and in particular samples with higher polydispersity have a higher $\phi_g$ \cite{pusey09,pednekar18}.  Thus for both of these reasons, our $\phi=0.56$ sample is likely closer to our $\phi_g$ than the prior work.  Near the colloidal glass transition, even modest increases in effective volume fraction can produce order of magnitude increases in the elastic modulus because of the strong dependence on proximity to dynamical arrest \cite{hunter12rpp}.  That being said, it is also plausible that differences could be due to our use of active microrheology rather than the macrorheology of Refs.~\cite{mason95,athanasiou25}, where our driven probe could be sensitive to local spatial heterogeneity.  Reassuringly, while the magnitude of the moduli differs in our experiment, the modest frequency dependence of our moduli data is similar to the literature results at the same $\omega \tau_0 = {\rm Pe}$.

\section{Colloidal Particles' Dynamics}

\begin{figure}
    \includegraphics[width=8cm,bb = 106 412 537 720]{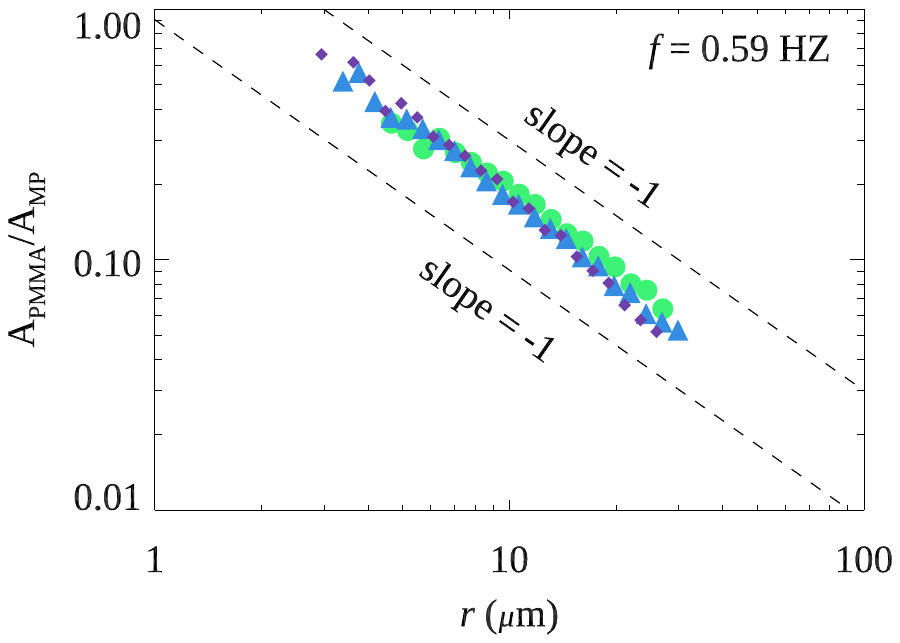}
    \caption{The amplitude of the colloidal particles, $A_{\rm PMMA}$, scaled by the amplitude of the probe particle, $A_{\rm MP}$, vs. distance from the center of the magnetic particle rotation, $r$ for selected volume fractions: $\phi =$ 0.32 (circles), 0.47 (triangles), and 0.56 (diamonds) and probe particle oscillation frequency 0.59 Hz.  For clarity, data for $\phi =$ 0.45, 0.52, and 0.55 are not shown.  The dashed lines have a slope indicating $1/r$ scaling}.
    \label{Apmmavsr}
\end{figure}

Since we use confocal microscopy in these experiments, it would be amiss if we did not take the advantage of the fact that we can calculate the trajectory of each colloidal particle in the field of view and also study the response of the colloidal suspension to the local perturbation at the single-particle level.

Figure~\ref{Apmmavsr} presents the average oscillatory amplitude of the colloidal particles scaled by the amplitude of the magnetic probe, $A_{\rm PMMA}/A_{\rm MP}$, as a function of their distance from the center of the magnetic particle, $r$, for a frequency of the magnetic probe of 0.59 Hz and selected volume fractions. Normalizing by the probe amplitude leads to an excellent collapse of the data across different driving forces, confirming that the propagation of displacements is governed by the material’s intrinsic mechanical response rather than by the absolute strength of the forcing.

The mean oscillatory amplitude of individual colloidal particles decays as $1/r$ from the probe particle across all volume fractions, matching theoretical predictions from continuum mechanics for the displacement field around an oscillating point force \cite{crocker00,levine00}.  The dashed reference lines with slope $-1$ on this log-log plot correspond to perfect $1/r$ scaling.  The $1/r$ scaling is significant for several fundamental reasons. First, it demonstrates that colloidal suspensions behave as effective continuum media even when probed on length scales of only a few particle diameters. This is somewhat surprising given the discrete, particulate nature of the material, since one might expect significant deviations from continuum predictions when individual particles are resolved. However, the excellent agreement with $1/r$ scaling arises because the cooperative motion of many particles produces an averaged response that appears continuous, validating the continuum approximation even at these microscopic scales \cite{schweizer03,schweizer07,anderson13}.  This is similar to a fluid, for which the continuum behavior is not seen at a molecular scale and short time scales, but for which the continuum behavior arises averaging over all molecules in a small region \cite{tritton88}.  In our colloidal experiments the continuum-like response is directly seen when averaging over the $O(10)$ particles touching the magnetic probe particle, and averaging over $O(100)$ oscillation periods.  In contrast to a prior experiment which studied a quasi-static elastic response \cite{anderson13}, here we see the continuum behavior in a dynamic experiment which includes the full viscoelastic sample response.

Interestingly, the $1/r$ decay is actually predicted for both purely viscous fluids and purely elastic solids in response to an oscillating point force, making it a universal feature of linear viscoelastic materials \cite{crocker00,levine00}.  This relates to the correspondence principle \cite{pipkin1986,furst17}, which notes that solutions to linear viscoelastic flow problems can be mapped directly onto solutions of a corresponding elasticity problem \cite{lee1955} or Stokes flow problem \cite{furst17}.  In the viscous limit, the Stokes flow solution for fluid displacement around a moving sphere gives $u(r) \sim F/(\eta r) \sim 1/r$ (\cite{crocker00}). In the elastic limit, the displacement field from a point force in an infinite elastic medium similarly decays as $u(r) \sim F/(G r) \sim 1/r$, where $G$ is the shear modulus of the infinite elastic medium \cite{crocker00,anderson13}. Our observation is that the same $1/r$ scaling holds across the full range of volume fractions: from viscous-dominated ($\phi$ = 0.32) to elastic-dominated ($\phi$ = 0.56), confirming that indeed our samples behave as continuum materials even at the particle length scale (when averaged).  Note that same $1/r$ scaling and data collapse was observed for other probe particle oscillatory frequencies.

Some scatter around the perfect $1/r$ behavior is evident, particularly at the highest volume fraction (diamonds in Fig.~\ref{Apmmavsr}). This likely reflects increasing structural heterogeneity and particle caging effects as the glass transition is approached. The range of measured distances (roughly 30 $\mu$m) is limited by our field of view and by the decreasing amplitude at large $r$, which eventually becomes comparable to thermal fluctuations and measurement uncertainty. Nevertheless, the reasonable $1/r$ scaling seen over this range provides strong validation of a continuum mechanics description.

In addition to the oscillation amplitude of the colloidal particles, the phase of the colloidal particle motion relative to the probe provides complementary insight into how stresses propagate through the suspension.  We expect the speed of sound in our colloidal suspensions to be set by the liquid and to be more than 1000 m/s \cite{ye93,riese99,bakker02}.  On the length scales we study, information is propagated via sound waves on time scales $O(10)$~ns, far faster than our frame rate.  Indeed, we measure the phase lag of the colloidal particles with respect to the magnetic particle, and find that it is essentially flat, confirming effectively instantaneous response of the colloidal particles to the nearby magnetic particle's motion.

\section{Conclusions}
We have investigated the local mechanical response of dense colloidal suspensions near the colloidal glass transition using magnetically driven probe particles in combination with confocal microscopy. This approach enables simultaneous measurement of probe dynamics and the displacement field of surrounding particles, linking microrheological response to spatially resolved particle motion.

The oscillatory motion of the magnetic probe shows a systematic change in response with increasing volume fraction $\phi$. At low $\phi$, the probe amplitude scales approximately as $1/\omega$ and the phase lag approaches $90^\circ$, consistent with viscous-dominated behavior. With increasing $\phi$, the frequency dependence weakens and the phase lag decreases, indicating an increasing elastic contribution associated with particle caging. From the probe amplitude and phase, we obtain the elastic and viscous moduli, $G'(\omega)$ and $G''(\omega)$. Both moduli increase with $\phi$, and their magnitudes and trends are consistent with conventional rheology at comparable volume fractions. At the highest $\phi$ studied, $G'(\omega) > G''(\omega)$ over the accessible frequency range, indicating predominantly elastic response.

Direct tracking of colloidal particles shows that the mean oscillatory displacement amplitude decays approximately as $1/r$ with distance $r$ from the probe. This scaling is consistent with the far-field response of a linear viscoelastic continuum to a localized oscillatory force. The observation of this behavior in both fluid-like and glassy samples indicates that continuum descriptions remain applicable down to length scales of only a few particle diameters, although increased scatter at high $\phi$ suggests growing structural heterogeneity.  The continuum behavior is consistent with the relatively small amplitude of the forced motion of the magnetic probes:  our prior work showed that in the linear regime, colloidal samples respond in agreement with continuum elasticity theory \cite{anderson13}, whereas for probes forced out of the linear regime, the sample responds via forced local rearrangements \cite{habdas04,habdas25}.  Because the probe radius is larger than the colloidal particle radius, the measured response represents the collective deformation of several neighboring particles. However, the small displacement amplitudes used here avoid cage breaking and irreversible rearrangements, allowing the response to be interpreted within linear viscoelasticity.

The phase of the particle motion does not exhibit a measurable dependence on distance from the probe within our resolution. This indicates that stress transmission across the observed length scales occurs on timescales short compared with the oscillation period, consistent with overdamped viscoelastic response.


\begin{acknowledgments}
We thank A. B. Schofield for synthesizing our colloidal particles.  We thank D. R. Nelson for the original inspiration for this project.  The initial data collection was supported by NASA (NAG3-2284) and subsequent data analysis was supported by the National Science Foundation (CBET-2002815, CBET-2333224).
\end{acknowledgments}

\section*{Data Availability}

The data that support the findings of this article are openly
available \cite{dataonline26}.


\end{document}